# Probing Optically Silent Superfluid Stripes in Cuprates

S. Rajasekaran[1], J. Okamoto[2], L. Mathey[2], M. Fechner[1], V. Thampy[3], G.D. Gu[3], A. Cavalleri[1,4]

[1]*Max Planck Institute for the Structure and Dynamics of Matter, Hamburg, Germany*
[2]*Centre for Optical Quantum Technologies and Institute for Laser physics, University of Hamburg, Germany*
[3]*Condensed Matter Physics and Materials Science Department, Brookhaven National Laboratory, USA*
[4]*Department of Physics, University of Oxford, Clarendon Laboratory, UK*

**Unconventional superconductivity in the cuprates coexists with other types of electronic order. However, some of these orders are invisible to most experimental probes because of their symmetry. For example, the possible existence of superfluid stripes is not easily validated with linear optics, because the stripe alignment causes interlayer superconducting tunneling to vanish on average. Here, we show that this frustration is removed in the nonlinear optical response. A giant Terahertz third harmonic, characteristic of nonlinear Josephson tunneling, is observed in $La_{1.885}Ba_{0.115}CuO_4$ above the transition temperature $T_c$=13 K and up to the charge ordering temperature $T_{co}$ = 55 K. We model these results by hypothesizing the presence of a pair density wave condensate, in which nonlinear mixing of optically-silent tunneling modes drive large dipole-carrying super-currents.**

Single-layer cuprates of the type $La_{2-x-y}(Ba,Sr)_x(Nd,Eu)_yCuO_4$ exhibit an anomalous suppression of the superconducting transition temperature $T_c$ for doping levels near 12.5% (*1*). Several studies have shown that this suppression coincides with the formation of "stripes", one-dimensional chains of charge rivers separated by regions of oppositely phased antiferromagnetism (*2*, *3*). A schematic phase diagram of $La_{2-x}Ba_xCuO_4$, adapted from ref (*1*), is shown in Fig. 1, along with a sketch of the stripe geometry.

Recently, high-mobility in-plane transport was found in some of these striped phases at temperatures above the bulk superconducting transition temperature $T_c$ (*4*). The existence of a striped superfluid state with a spatially oscillating superconducting order-parameter, a so-called Pair Density Wave state (PDW), was hypothesized (*5*) to explain the anomalously low in-plane resistivity.

Superfluid stripes are difficult to detect. Scanning Tunneling Microscopy (STM) experiments have reported spatial modulations in the superconducting condensate strength of $Bi_2Sr_2CaCu_2O_{8+x}$ (*6*). However, STM is not sensitive to the phase of the order parameter and these low temperature measurements did not clarify the broader question of whether finite momentum condensation may be taking place above $T_c$. Also, according to the pair density wave model, superfluid transport perpendicular to the planes is frustrated due to the stripe alignment (see sketch in figure 1). Hence, these stripes are invisible in *linear* c-axis optical measurements (*7*).

In this paper, we show that superfluid stripes display characteristic signatures in the nonlinear Terahertz (THz) frequency response. In $La_{1.885}Ba_{0.115}CuO_4$, we detect superfluid stripes above $T_c$=13 K and up to the charge ordering temperature $T_{CO}$=55 K.

Figure 2 displays the linear and nonlinear THz reflectivity spectra for two $La_{(2-x)}Ba_xCuO_4$ crystals (x=9.5% and x=15.5%), measured with single-cycle pulses (*8*) polarized along the c-axis (see section S2 of (*9*)) and covering the spectral range between 150 GHz and 2.5 THz (see Section S2 of (*9*)). At these doping levels, the material exhibits homogenous superconductivity and only weak stripe order, with transition temperatures of $T_{co} \sim T_c = 34$ K (x=9.5%) and $T_{co} \sim 40$ K $> T_c$=32 K (x=15.5%).

The low temperature linear reflectivities (T = 5 K $<$ $T_C$), show Josephson plasma edges at $\omega_{JP0}$ = 500 GHz and $\omega_{JP0}$ = 1.4 THz for x=9.5% and x=15.5%, respectively (see dashed curves in Figs. 2, A and C). The reflectivity edges shifted to lower frequencies with increasing temperature, indicating a decrease in superfluid density (Figs. 2, B and D, section S2 of (*9*)).

Nonlinear reflectivities, measured at field strengths between 20 kV/cm and 80 kV/cm (see Figs. 2, A and C) displayed two characteristic effects of nonlinear interlayer Josephson coupling: (i) a field-dependent red-shift of the plasma edge (*10–12*), and (ii) a reflectivity peak at the third harmonic of the pump field $\omega \approx 3\omega_{pump}$(*13*). The third harmonic field amplitude $E(3\omega_{pump})$ scaled with the cube of the incident field strength $E(\omega_{pump})$ (Fig. 2E) and decreased in strength with temperature and disappeared for T=$T_c$, tracking the superfluid density $\omega\sigma_2(\omega \to 0)$ (see Figs. 2, C, D and F).

These observations are well understood by a semi-quantitative analysis of the interlayer phase dynamics of a homogeneous layered superconductor (*10*, *12*, *14*). For a c-axis electric field $E(t) = E_0 \sin(\omega_{pump}t)$ the interlayer phase difference $\theta(t)$ advances in time according to the second Josephson equation $\frac{\partial \theta(t)}{\partial t} = \frac{2edE(t)}{\hbar}$, where d is the interlayer spacing ($\sim$ 1nm), 2e is the Cooper pair charge and $\hbar$ is Planck's constant (*15*). Because the c-axis superfluid density $\rho_c$ scales with the order parameter phase difference $\rho_c \propto \cos(\theta)$ and because $\rho_c \propto \omega_J^2$, the plasma frequency renormalizes as $\omega_J^2 = \omega_{JP0}^2 \cos(\theta) = \omega_{JP0}^2 \cos(\theta_0 \cos(\omega_{pump}t)) \approx \omega_{JP0}^2 \left(1 - \frac{\theta_0^2}{4} - \frac{\theta_0^2 \cos(2\omega_{pump}t)}{4}\right)$, where $\theta_0 = 2edE_0/\hbar\omega_{pump}$, Hence, an average redshift of the equilibrium plasma resonance $\omega_{JP0}$ is estimated as $\omega_J^2 = \omega_{JP0}^2 \left(1 - \frac{\theta_0^2}{4}\right)$. Secondly, a tunneling supercurrent is excited at the first and third harmonic of the driving field, $I(t) =$

$I_c \sin(\theta_0 \cos(\omega_{pump} t)) \approx I_c \left(\theta_0 \cos(\omega_{pump} t) - \frac{\theta_0^3}{6} \cos^3(\omega_{pump} t)\right)$, giving rise to third harmonic radiation. Finally, because the third harmonic signal is proportional to $I_c$, it is expected to follow the same temperature dependence as the superfluid density.

More comprehensive numerical simulations, based on space (x) and time (t) dependent one-dimensional sine-Gordon equation for the Josephson phase $\theta(x,t)$ (see supplementary Section S3 of (9)) (12, 14, 16, 17) were used to obtain the electromagnetic field at the surface of the superconductor and to calculate the reflectivity for arbitrary field strengths. These simulations, reported in Figure 3, reproduce the experimental data closely.

We next turn to the key results of this paper, which provide evidence for superfluidity in the normal-state striped phase in $La_{1.885}Ba_{0.115}CuO_4$ (x=11.5%). In this compound, striped charge order coexists with superconductivity below $T_c$ = 13 K and extends into the normal state up to $T_{co}$=55 K (1).

Fig. 4A displays the results for T < $T_c$ = 13 K. Note that in this material, the equilibrium Josephson plasma resonance was at lower frequency than in the other two compounds with higher $T_c$ (~150 GHz) and could not be observed (see Section S2 of (9)). However, a third harmonic at $\omega = 3\omega_{pump} \approx 1.4\ THz$ was clearly observed. Strikingly, the third harmonic signal remained finite also for T > $T_c$ = 13 K and up to T~$T_{co}$ =55 K (Fig. 4, C and D). A giant third harmonic, amounting to several percent of the driving field, can be understood only in presence of superconducting tunneling above $T_c$ and up to $T_{co}$.

In the following, we show that a pair density wave, which does not show features of superfluidity in the linear optical properties, retains the large nonlinear optical signal of the homogeneous condensate. As shown in Fig. 4E, the order parameter phase of the PDW, encoded by the vector angle (black arrow), changes between neighboring stripes and rotates by 90 degrees from one plane to the next (18), resulting in a checkerboard lattice of $\frac{\pi}{2}$ and $-\frac{\pi}{2}$

Josephson Junctions. Hence, the equilibrium PDW supports a lattice of staggered tunneling supercurrents, which average out to zero (thick red arrows in Fig. 4E).

The interlayer phase fluctuations probed in the optical response are described by two normal modes, termed here $\varphi_0$ and $\varphi_\pi$. The $\varphi_0$ mode is optically active. For an optical field $E(t) = E_0 \sin(\omega_{pump} t)$, one has identical phase excursions $\delta\theta_0(t) = \frac{2edE_0}{\hbar\omega_{pump}} \cos(\omega_{pump} t)$ at each site (shading under black arrows in fig. 4F). However, current fluctuations $\delta I_0(t)$ of equal magnitude but opposite sign at neighboring $\frac{\pi}{2}$ and $-\frac{\pi}{2}$ junctions, makes this mode silent (red arrows in fig. 4F). The second $\varphi_\pi$ mode consists of phase excursions $\delta\theta_\pi(t)$ occurring in opposite directions at neighboring sites, and is optically inactive (see fig. 4F).

In the nonlinear regime the optical response of the PDW is no longer zero. Because $\varphi_0$ is odd and $\varphi_\pi$ is even, a nonlinear expansion of the Josephson energy can be written as $U(\varphi_0, \varphi_\pi) \propto J_0 \varphi_\pi^2 - J'' \varphi_\pi + C \varphi_0^2 \varphi_\pi$, where $J_0$ and $J''$ are the in-plane and out-of-plane Josephson energies and $C = J''/2$, with $J_0 \gg J''$ (see Section S5 of (9)).

A large third harmonic signal is readily predicted from the $\varphi_0^2 \varphi_\pi$ coupling term. For an optical field of the type $E(t) = E_0 \sin(\omega_{pump} t)$, which acts only on the mode $\varphi_0$, the equations of motion of the phase are $\ddot{\varphi}_0 \approx \omega_{pump} E_0 \cos(\omega_{pump} t)$ and $\ddot{\varphi}_\pi \approx -2C\varphi_0^2$. These coupled equations imply not only phase oscillations in $\varphi_0$ mode at $\omega_{pump}$, $\delta\theta_0(\omega_{pump})$, but also indirect excitation of the optically inactive $\varphi_\pi$ mode. Because in the equation of motion for $\varphi_\pi$ the driving force is proportional to $\varphi_0^2$, the phase of this mode $\delta\theta_\pi$ is driven at $2\omega_{pump}$. Because the total nonlinear current $I_{tot} = \sum I_{junctions}$ contains terms of the type $I_{tot}^* \sim 2I_c \varphi_0 \varphi_\pi$, the two phase coordinates $\varphi_0(\omega_{pump})$ and $\varphi_\pi(2\omega_{pump})$ are mixed, and produce current oscillations at the difference and sum frequencies $\omega_{pump}$ and $3\omega_{pump}$. A numerical solution of the two equations of motion with realistic parameters for the Josephson coupling energies $J_0$ and $J''$, displays oscillatory currents at the fundamental and third harmonic (Fig. 4G, further see Section S5 of (9) for details on the simulations).

Finally, from the model above the third harmonic current is predicted to scale linearly with the out-of-plane critical current, and hence the superfluid density of the stripes. From the plot in figure 4D, the local superfluid density at $T_{so}$ is found to be 60% of that measured below $T_c$, and subsequently decreases continuously as the temperature is increased further, before vanishing at $T_{co}$.

Although the model discussed above provides a plausible description of the experimental observations, other hypotheses for the origin of the third harmonic signal for $T > T_c$ should be considered. The measured and simulated third harmonic is far larger than, and hence easily distinguished from, the effect of non-condensed quasi-particles. The nonlinear susceptibility detected in the present experiments $\chi^{(3)} \sim 10^{-15}\ m^2/V^2$ (obtained from the electric field strengths at the fundamental and third harmonic), is several orders of magnitude bigger than typical cubic nonlinearities ($\chi^{(3)} \sim 10^{-18} - 10^{-20} m^2/V^2$)(*19, 20, 21*). Furthermore, first principle calculations show that for this compound the value of $\chi^{(3)}$ from quasi-particle transport within anharmonic bands is at least three orders of magnitude smaller than what is measured here (Section 6 of (*9*)).

A second alternative may involve the sliding of a Charge Density Wave along the *c* axis, as discussed in ref. (*22*) for Blue Bronze. However, the efficiency of sliding of a charge density wave, reported in reference (*22*) for KHz frequency excitation, is expected to reduce strongly at higher excitation frequencies and can be ruled out for the THz irradiation (see section S7 of (*9*)). The results on the 15.5% sample, where the third harmonic signal disappears at $T_c$ = 32 K (< $T_{CO}$ = 40 K) further indicate that the third harmonic results from superconducting tunneling rather than charge ordering.

The observation of a colossal third harmonic signal in the stripe ordered state of $La_{1.885}Sr_{0.115}CuO_4$ provides compelling experimental evidence for finite momentum condensation in the normal state of cuprates, and underscore the power of nonlinear Terahertz optics as a sensitive probe of frustrated excitations in quantum solids. A natural direction for this line of research involves the study of other forms of charge order that compete or coexist with superconductivity such as those found in $YBa_2Cu_3O_{6+x}$(*23*, *24*). One may also find application of these techniques in other regimes of the cuprate pseudogap, with finite superfluid density, vanishing range phase correlations(*25*, *26*) or where other forms of density waves(*27*, *28*) have been discussed.

**Acknowledgments**

We thank S.A. Kivelson and J.M. Tranquada for helpful feedback on the manuscript. We thank D. Nicoletti for providing us the linear reflectivity data of the x=11.5% sample. The research leading to these results received funding from the European Research Council under the European Union's Seventh Framework Programme (FP7/2007-2013)/ERC Grant Agreement no. 319286 (QMAC). Work performed at Brookhaven was supported by US Department of Energy, Division of Materials Science under contract no. DE-AC02-98CH10886. The data from the THz measurements and the simulations are kept at the Cavalleri Laboratory at Max Planck Institute of Structure and Dynamics, Hamburg.

**Supplementary Materials**

Materials and Methods

Supplementary Text

Figures S1 to S8

Equations eq. S1 to eq. S15

References (29-34)

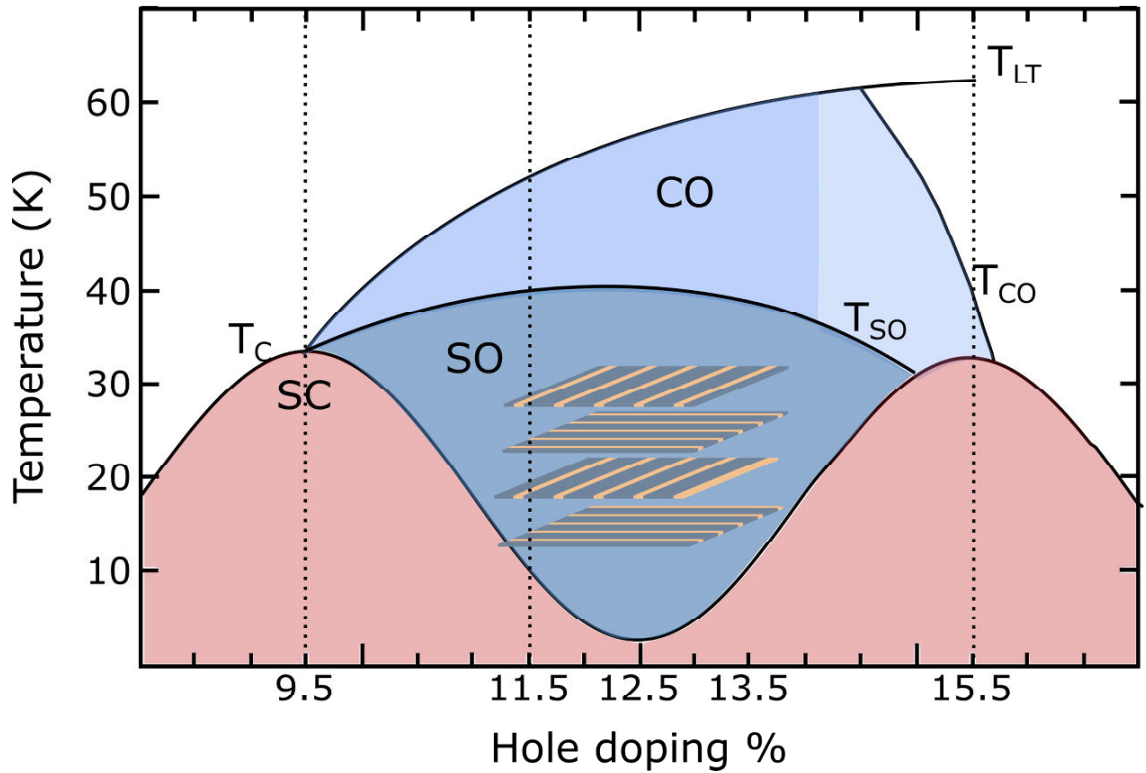

**Figure 1: Phase diagram for $La_{2-x}Ba_xCuO_4$.** Adapted with permission from Ref. (1). SC, SO and CO denote the bulk superconducting, spin and charge ordered (striped), and charge-only ordered phases, respectively. $T_c$, $T_{so}$ and $T_{co}$ are the corresponding ordering temperatures. $T_{LT}$ denotes the orthorhombic to tetragonal structural transition temperature. The samples examined in this study are x = 9.5 %, x=11.5% and x = 15.5 % (dotted lines). Further, a schematic stripe ordered state is shown wherein the yellow stripes depict the charge rivers and the grey stripes, the antiferromagnetic insulating region.

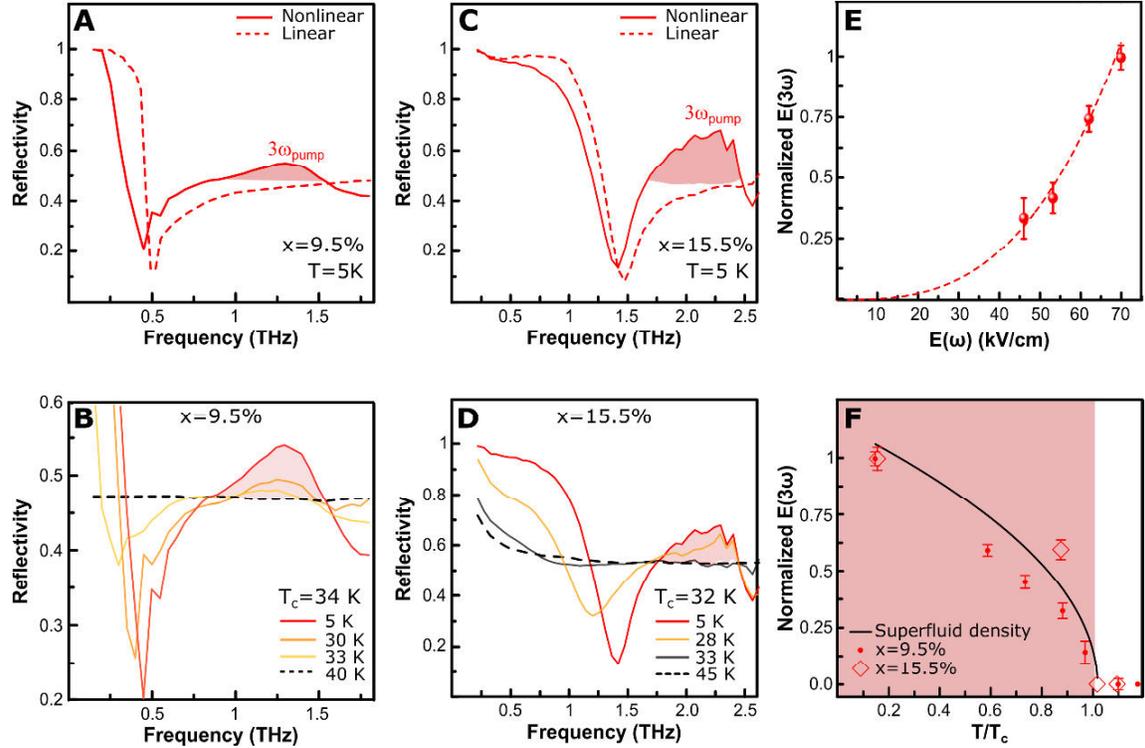

**Figure 2: Third Harmonic from homogeneous superconductor.** (A) Linear and nonlinear reflectivity of $La_{2-x}Ba_xCuO_4$ x=9.5% measured at T=5 K with $\omega_{pump}=450 GHz$. The linear reflectivity displays a Josephson plasma edge at $\omega_{JP0}$ = 500 GHz (0.5 THz), whereas the nonlinear reflectivity shows a red shift of the edge and third Harmonic Generation (THG-red shading). (B) Temperature dependence of nonlinear reflectivity for x=9.5%. The third harmonic peak disappears above $T_c$=34 K. (C) Linear and nonlinear reflectivity for x=15.5%, measured with $\omega_{pump}=700 GHz$. (D) Temperature dependence of nonlinear reflectivity for x=9.5%. The third harmonic peak disappears above $T_c$=34 K. Third harmonic generation (red shading), which disappears above $T_c$=32 K. (E) Third harmonic electric field strength (normalized to the highest signal) plotted as a function of the incident electric field strength (defined as explained in Section S1 of (9)) measured at T=5 K from the x=9.5% sample. The third harmonic field displays a cubic dependence on the incident field strength. (F) Temperature dependence of the third harmonic amplitude (normalized to the measurement at

T= 5 K) from the x=9.5% and 15.5% doping. The superfluid density ($\omega\sigma_2(\sigma \to 0)$) (normalized to the measurement at T= 5 K), extracted from the linear optical properties of the x=9.5% sample is also shown. All the quantities vanish above $T_c$.

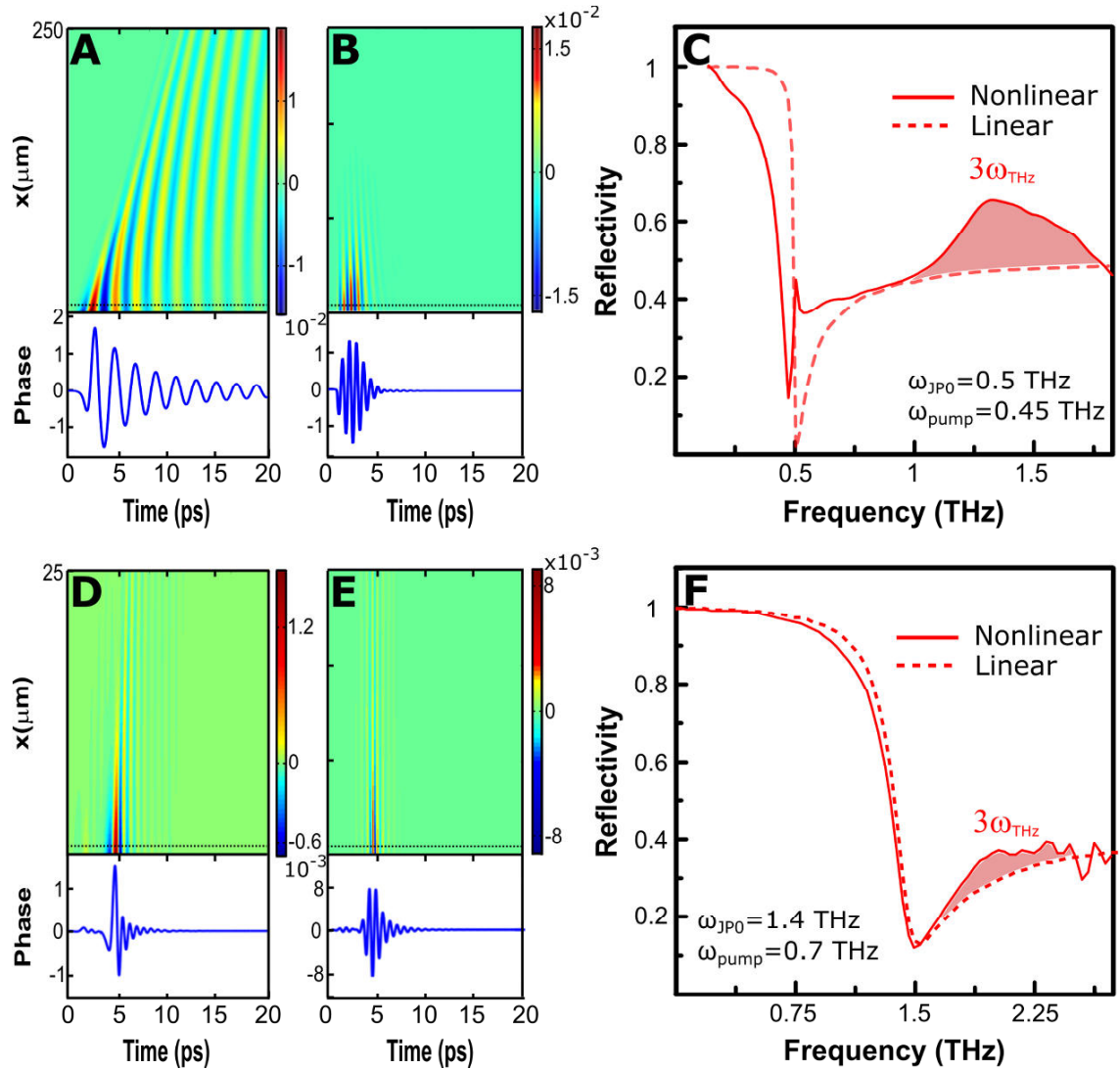

**Figure 3: Simulated nonlinear reflectivity for homogeneous superconductivity.** Simulations at x=9.5% and x=15.5% doping. (A) Simulated space and time dependent order parameter phase ($\theta(x,t)$), obtained by numerically solving the Sine-Gordon Equation on x=9.5% samples (See Section S3 of (9)). The equation makes use of equilibrium superfluid density extracted from the linear optical properties and assumes excitation with Terahertz pulses of shape and strength used in the experiment. The horizontal dotted lines indicate the spatial coordinate $x$ at which the line cuts are displayed (lower panels). (B) Simulated order parameter phase (panel A) after frequency filtering centered at $3\omega_{pump}$ with its corresponding

line cut (lower panel) (C) Simulated reflectivity in the linear ($E = 0.1\ kV/cm$) and the nonlinear ($E = 80\ kV/cm$) regime. The THG component is highlighted (red shading). (D-F) As in (A-C) but for x=15.5% .

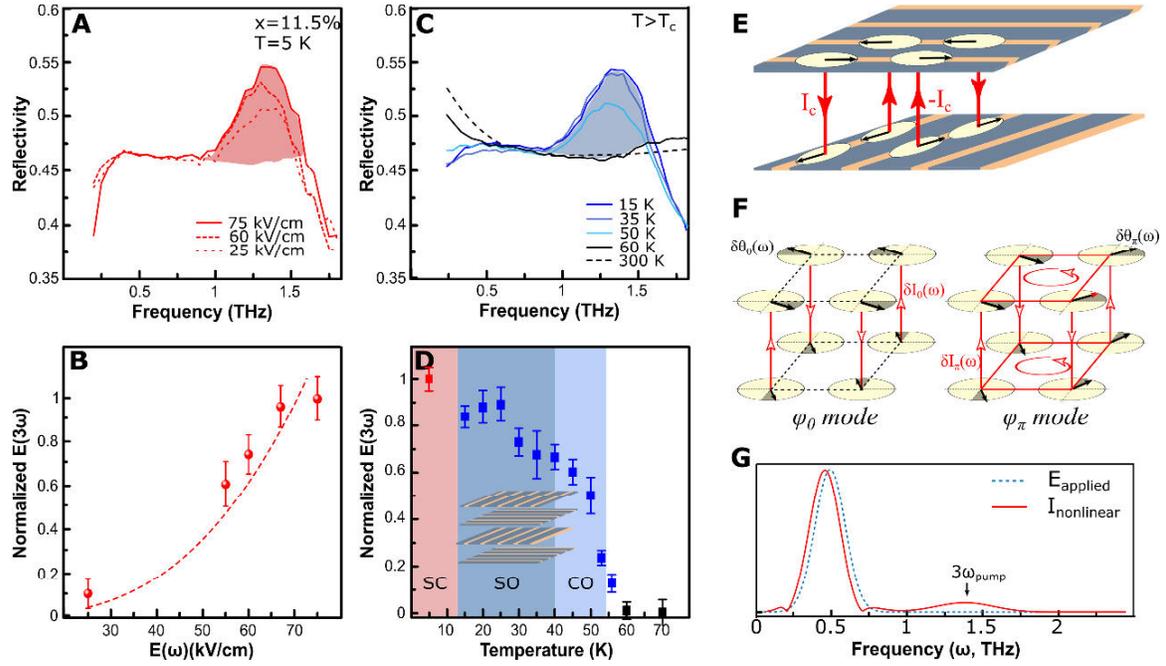

**Figure 4: Third Harmonic generation in the striped phase.** (A) Nonlinear frequency dependent reflectivity measured in the striped x=11.5% samples recorded for three different field strengths at T =5 K (< $T_c$ = 13 K). (B) Electric - field dependence of the third harmonic amplitude for T = 5 K. (C) Temperature dependence of the nonlinear reflectivity for T > $T_c$ = 13 K. (D) Temperature dependence of the third harmonic signal (normalized to the highest field measurements at T= 5 K). (E) Schematic of the order-parameter phase in a Pair Density Wave condensate. The black arrows represent the superconducting order parameter phase at each lattice point. Interlayer tunneling from perpendicularly aligned superfluid stripes is equivalent to a checkerboard lattice of alternating π/2 and – π /2 phase Josephson junctions. Such a lattice has tunneling currents of $I_c$ and $–I_c$ flowing at the neighboring junctions at equilibrium (thick red lines). (F) Excitation modes of the Pair Density Wave state indicating the $\varphi_0$ and $\varphi_\pi$ modes (see text). The shaded region under the black arrow represents the phase excursion from the equilibrium geometry ($\delta\theta_0$ and $\delta\theta_\pi$). Corresponding current fluctuations $\delta I_0$ and $\delta I_\pi$ produced by such excitations are also depicted (thin red lines). (G) Calculated

nonlinear current response for the unit cell of (E) after application of a single cycle optical pulse centered at 500 GHz frequency (see Section S5 of (9) for details).

# Probing Optically Silent Superfluid Stripes in Cuprates

S. Rajasekaran[1], J. Okamoto[2], L. Mathey[2], M. Fechner[1], V. Thampy[3], G.D. Gu[3], A. Cavalleri[1,4]

[1]*Max Planck Institute for the Structure and Dynamics of Matter, Hamburg, Germany*
[2]*Zentrum für Optische Quantentechnologien and Institut für Laserphysik, Universität Hamburg, Germany*
[3]*Condensed Matter Physics and Materials Science Department, Brookhaven National Laboratory, USA*
[4]*Department of Physics, University of Oxford, Clarendon Laboratory, UK*

## Supplementary Material

## S1: Methods

Large single crystals of $La_{2-x}Ba_xCuO_4$ with x=9.5%, 11.5% and 15.5% (~4 mm diameter), grown by transient solvent method, were studied here. These crystals belonged to the same batch of samples as reported in an earlier work (*7*), and were cut and polished along the *ac* surface.

Laser pulses at 800-nm wavelength, 100-fs duration and 5 mJ energy were split into 2 parts (99%, 1%) with a beam splitter. The most intense beam was used to generate Terahertz (THz) pulses by optical rectification in $LiNbO_3$ with the tilted pulse front technique(*8*). These THz pulses had energies of ~3 µJ. The pulses were collimated and focused at an incidence angle of 20° onto the samples. The THz pulses were *s*-polarized (i.e., perpendicular to the plane of incidence), corresponding to the direction perpendicular to the Cu-O planes (parallel to the *c* axis, see Fig. S1).

The THz beam spot diameter at the sample position was 2.5 mm, corresponding to a maximum attainable field strength of ~80 kV/cm. The incident field strength was adjusted using a pair of wire grid polarizers.

After reflection from the sample surface, the THz pulse was then electro-optically sampled in a 0.2-mm-thick GaP crystal using the 1% fraction of the 800-nm beam.

This measurement procedure returned the quantity $E_{reflected,sample}(t)$, with *t* being the electro-optic internal time delay. The incident field was measured after reflection from a gold reference, i.e. $E_{reflected,gold}(t)$. The frequency-dependent reflectivity *R* was then derived after computing the Fourier transforms of the time domain THz fields as $R = |r|^2 = |E_{reflected,sample}(\omega)/E_{reflected,gold}(\omega)|^2$.

Linear reflectivities were recorded at the lowest achievable field strengths while nonlinear reflectivities were taken at higher field strengths (20 kV/cm < E < 80 kV/cm). Incident pulses with central frequency of $\omega_{pump}$=0.45 THz was used for the measurements in x=9.5% and 11.5% doping. The "weight" of the third harmonic component in the nonlinear reflectivity (800 GHz and 1.5 THz, red and blue shaded regions in figure 2 and 4 in main text), therefore could be obtained by subtracting normalized linear reflected electric field (with gold) taken at low fields from the normalized nonlinear electric field taken at high fields (i.e. the integrated $r_{nonlinear} - r_{linear} = \frac{E_{sample,nonlinear}}{E_{gold,nonlinear}} - \frac{E_{sample,linear}}{E_{gold,linear}}$ in the frequency range between 800 GHz and 1.5 THz). Note that for the x=15.5% sample in order to obtain a clear third harmonic signal without interference from the reflectivity edge at the plasma frequency (1.4 THz), THz pulses with $\omega_{pump}$=0.7 THz, was used. Hence for this case, a similar procedure explained earlier was used, albeit for the frequency range 1.7 THz-2.3 THz. The THz field transient and the corresponding spectrum was to obtain the measurements are shown in Section S3.

Equivalent results were obtained by subtracting the reflected *high field* signal at high temperatures (field-independent) from that measured in the reported ranges (field-dependent).(i.e. the integrated $E_{sample,T} - E_{sample,T\gg T_{CO}} \propto r_T - r_{T\gg T_{CO}}$, where $r_{T\gg T_{CO}}$ is field-independent) .

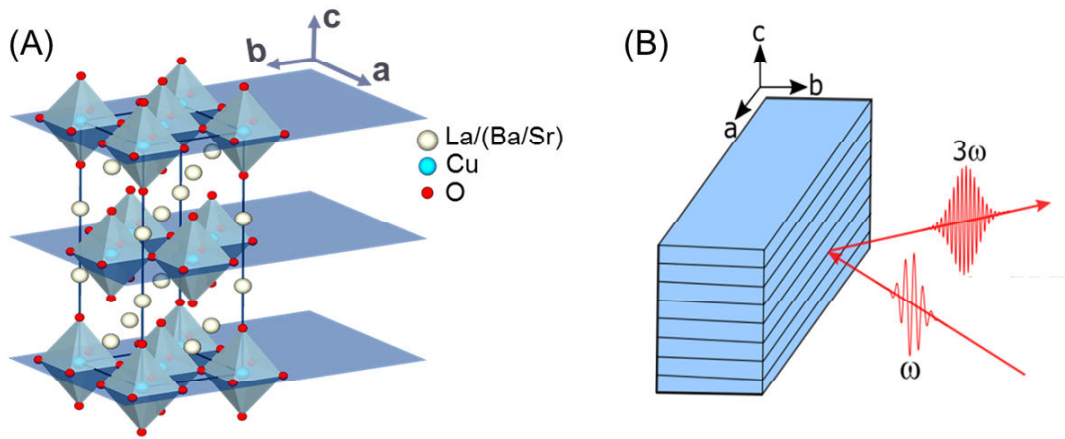

Figure S1 – (A) Schematic crystal structure of $La_{2-x}Ba_xCuO_4$ indicating the *c* axis stacked Cu-O planes. (B) Schematic representation of the experimental geometry.

## S2: Linear optical properties of $La_{2-x}Ba_xCuO_4$ - x=9.5% & x=11.5%

As shown in Fig. S2, the linear reflectivity of the x=9.5% sample shown in Fig. S1 displays a temperature dependent red shift of the Josephson plasma resonance.

The corresponding imaginary part of the optical conductivity, $\sigma_2(\omega)$, is also displayed. This was determined by Kramers-Kronig transformation, after merging the THz frequency reflectivity data reported here with the high frequency spectra reported in the literature for the same batch of samples(7). The superfluid density $\omega\sigma_2(\sigma \to 0)$ plotted in Fig. 2 was computed from the data shown here.

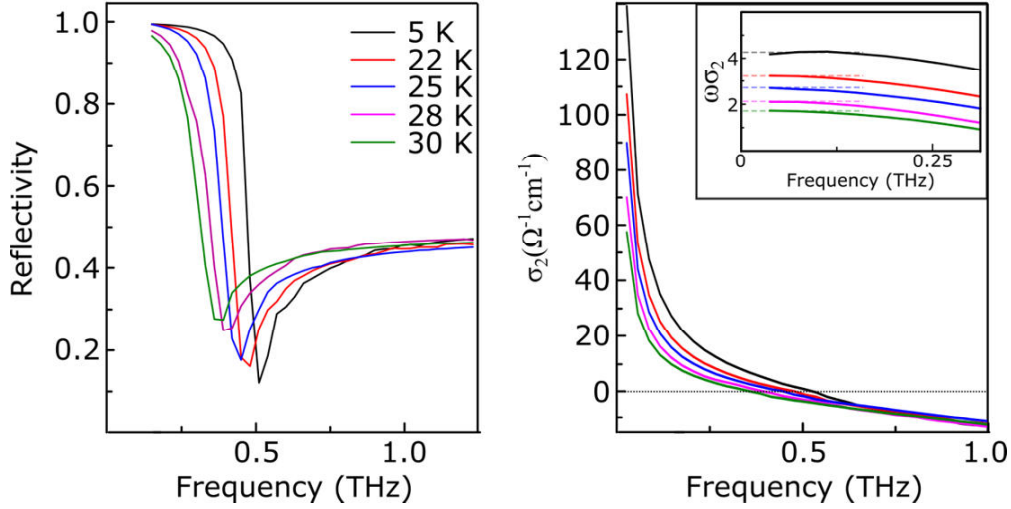

Figure S2 – (A) Temperature dependent reflectivity of x=9.5% doped sample measured in the linear regime (i.e. E=8kV/cm) (B) The corresponding $\sigma_2$ extracted through a Kramers-Kronig transformation. The superfluid density $\omega\sigma_2(\sigma \to 0)$ is shown in the inset.

The THz reflectivity of the x=11.5% sample in the linear regime, measured in a different experimental setup (based on a photoconductive antenna for THz generation) is shown in Fig. S3. We find indications of a Josephson Plasma edge at approximately 150 GHz(*29*). This feature is observed to disappear as T approaches $T_c$=13 K.

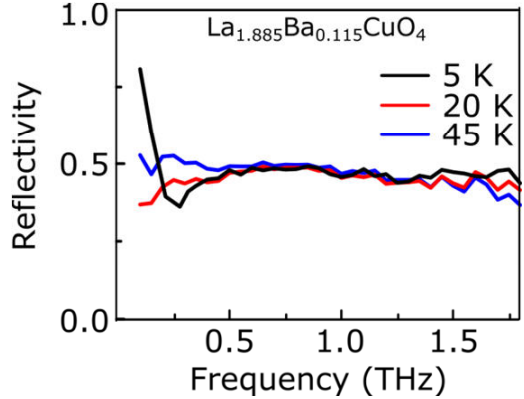

Figure S3: THz reflectivity in the linear regime of x=11.5% sample.

## S3: Simulation of the nonlinear optical properties from the sine-Gordon equation

A Josephson junction with semi-infinite layers stacked along the *z* direction (with translational invariance along the *y* direction) was modeled with the one-dimensional sine-Gordon equation(*16, 17*). The Josephson phase evolution in each stack $\theta_{i,i+1}(x,t)$, (with x being the propagation direction) is described by:

$$\frac{\partial^2 \theta_{i,i+1}(x,t)}{\partial x^2} - \frac{1}{\gamma}\frac{\partial \theta_{i,i+1}(x,t)}{\partial t} - \frac{\varepsilon_r}{c^2}\frac{\partial^2 \theta_{i,i+1}(x,t)}{\partial t^2} = \frac{\omega_{JP0}^2 \varepsilon_r}{c^2}\sin\theta_{i,i+1}(x,t) \quad \text{(S1)}$$

The damping factor $\gamma$ is a fitting parameter used to reproduce the experimental observations. For simplicity, i.e. we redefine $\theta_{i,i+1}(x,t) = \theta(x,t)$.

The Josephson phase evolution and the reflected field ($E_r$) are computed through the sine-Gordon equation along with the following boundary conditions at the vacuum-sample interface(*12, 14*).

$$[E_i(t) + E_r(t)]_{x=-0} = E_c(x,t)|_{x=+0} = H_0 \frac{1}{\omega_{JP0}\sqrt{\varepsilon}}\frac{\partial \theta(x,t)}{\partial t}\bigg|_{x=+0}, \quad \text{(S2)}$$

$$[H_i(t) + H_r(t)]_{x=-0} = H_c(x,t)|_{x=+0} = -H_0 \lambda_J \frac{\partial \theta(x,t)}{\partial x}\bigg|_{x=+0}. \quad \text{(S3)}$$

Here $E_c$ denotes the field propagating inside the superconducting cuprate, $H_0 = \Phi_0/2\pi D\lambda_J$, where $\Phi_0$ is the flux quantum $\left(\Phi_0 = \frac{hc}{2e}\right)$, $\lambda_J$ is the field penetration depth and D is the distance between adjacent superconducting layers. The equilibrium Josephson Plasma Resonance is an input parameter in the simulations, which is chosen to be that measured in linear spectrum, i.e. $\omega_{JP0} = 0.5$ THz and $\omega_{JP0} = 1.4$ THz for x=9.5% and 15.5% doping, respectively.

The THz field impinging ($E_i$) on the superconductor at the boundary $x = 0$ was taken as the digitized experimental field reflected from the gold reference, $E_{\text{gold}}$ (Fig. S4). For fields in vacuum ($x < 0$), the Maxwell's equations imply

$$E_i - E_r = \frac{\omega\mu}{ck}(H_i + H_r) = H_i + H_r. \quad \text{(S4)}$$

By combining Eq. (S3) with Eq. (S1) and (S2) we obtain the boundary condition

$$\frac{2\sqrt{\varepsilon}}{H_0}E_i(t)|_{x=-0} = \frac{\partial \theta(x,t)}{\omega_{JPR}\partial t}\bigg|_{x=+0} - \sqrt{\varepsilon}\frac{\partial \theta(x,t)}{\partial x/\lambda_J}\bigg|_{x=+0}. \quad \text{(S5)}$$

After solving the Josephson phase through Eq. (S1) and Eq. (S5), the reflected field is obtained from Eq. (S4). The reflectivity is then computed as the ratio between the Fourier transforms of the reflected field and the input field as

$$R(\omega) = |E_r(\omega)/E_i(\omega)|^2.$$

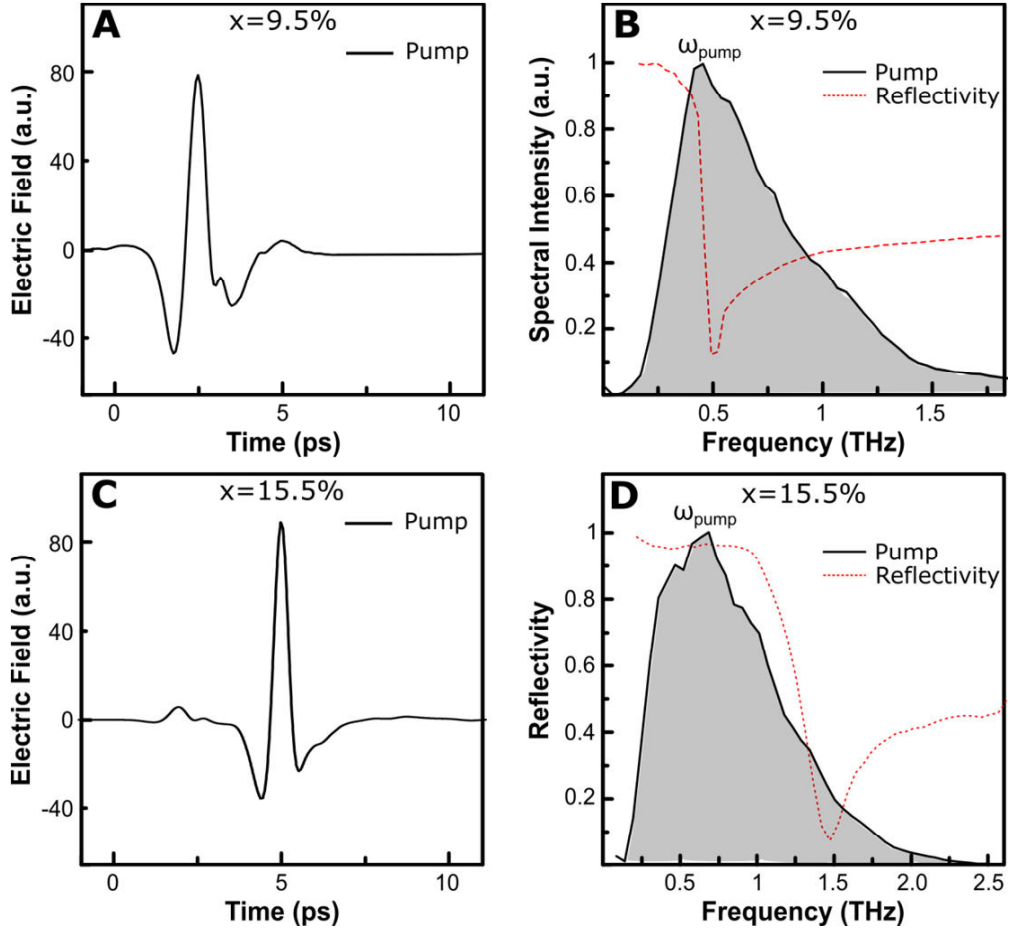

Figure S4 – (A) The incident THz field in time domain and (B) its frequency spectrum for the experiments on x=9.5% doping. The linear reflectivity of the sample at T=5 K is also shown. (C-D) The corresponding THz fields used for x=15.5% doping.

## S4: Third Harmonic generation with frequency filtered pulses

The nature of the third harmonic generation in the bulk superconducting state was further investigated using THz pulses that were filtered to less than 10%. Intense half cycle THz pulses, generated from optical rectification in LiNbO$_3$ using tilted pulse front technique (*8*), were shaped to narrow band multi-cycle pulses, with central frequency of $\omega_{pump}$~0.5 THz, by utilizing commercial frequency filters. These pulses had maximum field strengths of ~15 kV/cm and could therefore only be used to repeat a fraction of the experiments reported in the main text.

In Fig. S5A we show the frequency spectrum of the incident and the reflected THz pulses from the x=9.5% sample at T=5 K. In the reflected spectrum, a dip at 0.5 THz is observed, which corresponds to the Josephson Plasma Edge. In addition, a peak is observed at ~1.4 THz (inset), which corresponds to the third harmonic of the incident frequency. This feature is clearly captured by dividing the Fourier transform of the reflected field by the incident field (Fig. S5B).

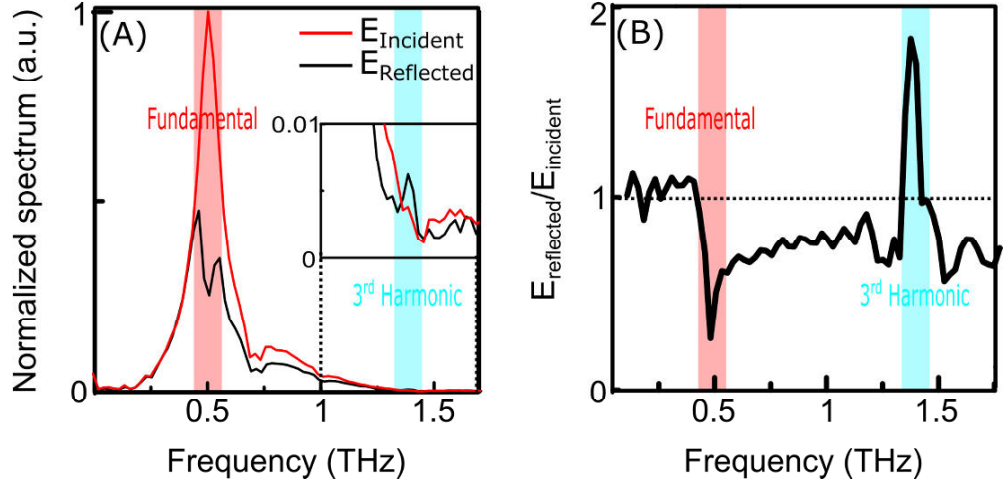

Figure S5 – (A) Fourier transforms of the incident and reflected THz fields from the x=9.5% sample at T=5 K. The inset shows a zoom in the 1 THz -1.8 THz frequency range. (B) Corresponding ratio $E_{reflected}/E_{incident}$ evidencing a third harmonic peak at ~1.4 THz.

## S5: Nonlinear response of Pair Density Wave – Two π/2 Josephson Junction model

In this section, we describe the two-site model which was utilized to compute the nonlinear response of the pair density wave (PDW). Such a PDW state could be modeled as a three dimensional array of Josephson junctions(*18*) (represented in Fig. S6). The total energy of such Josephson junctions with the intra-stripe, inter-stripe, and inter-layer couplings $J$, $J'$ and $J''$ could be written as

$$E_{total} = \sum_{x,y} -J\cos(\theta^1_{x,y} - \theta^1_{x,y+1}) + J'\cos(\theta^1_{x,y} - \theta^1_{x+1,y}) - J\cos(\theta^2_{x,y} - \theta^2_{x+1,y}) + J'\cos(\theta^2_{x,y} - \theta^2_{x,y+1}) - J''\cos(\theta^1_{x,y} - \theta^2_{x,y}) - Un^1_{x,y}n^2_{x,y},$$

where $\theta^1_{x,y}$ and $n^1_{x,y}$ are the superconducting phase and the charge at site $(x,y)$ in $i^{th}$ layer and $U$ is the inter-layer capacitive energy. Usually $J \sim J' \gg J''$ and for the case $J'' = 0$, the ground state minimizing this energy is indeed a stripe phase configuration with $\widetilde{\theta^1_{x,y}} = \left(x + \frac{1}{2}\right)\pi$ and $\widetilde{\theta^2_{x,y}} = y\pi$.

In the following we assume $J = J' = J_0$ and for simplicity we consider two $\pi/2$ Josephson Junctions at sites $r = 1, 2$ (four lattice points). In order to study the fluctuation dynamics induced by the external driving $E(t)$, we expand the phases at each lattice point around the stripe order as $\theta_r^i = \widetilde{\theta_r^i} + \delta\theta_r^i$ where $i$ represents the layer index. The energy of the two site model is therefore:

$$E_{two-site} = -J\cos(\delta\theta_1^1 - \delta\theta_2^1) - J\cos(\delta\theta_1^2 - \delta\theta_2^2) - \sum_{r=1,2}[Un_r^1 n_r^2 - (-1)^r J''\sin(\delta\theta_r^1 - \delta\theta_r^2)].$$

We emphasize the unusual nature of the inter-layer Josephson coupling energy [the term '$J''\sin(\delta\theta_r^1 - \delta\theta_r^2)$'], which arises due to the frustration nature of the PDW, is crucial towards observing the third harmonic nonlinearity.

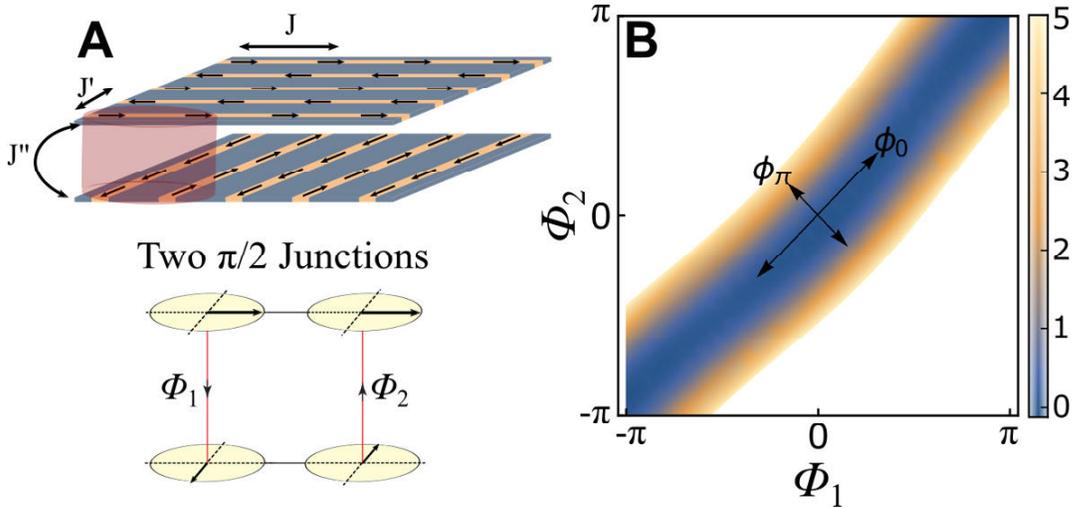

Fig. S6 – (A) Schematic representation of the Pair Density Wave and the corresponding simplified version involving two $\pi/2$ Josephson Junctions. (B) The effective potential ($V_{eff}$) used to simulate the nonlinear dynamics $J = 4, J'' = 0.5$ and $U = 1$.

The equation of motion of the inter-layer phase difference $\varphi_r = \delta\theta_r^1 - \delta\theta_r^2$ and its conjugate variable $(n_r^1 - n_r^2)$ are

$$\dot\varphi_r = \frac{\delta E_{two\,site}}{\delta n_r^1} - \frac{\delta E_{two\,site}}{\delta n_r^2} = U(n_r^1 - n_r^2),$$

$$\dot n_r^1 - \dot n_r^2 = -\frac{\delta E_{two-site}}{\delta(\delta\theta_r^1)} + \frac{\delta E_{two-site}}{\delta(\delta\theta_r^2)} = (-1)^r J(\varphi_1 - \varphi_2) - (-1)^r 2J''\cos(\varphi_r),$$

where the approximation $\sin(\delta\theta_1^i - \delta\theta_2^i) \approx (\delta\theta_1^i - \delta\theta_2^i)$ was made, since the in-plane fluctuations are small due to large $J$. The equations of motion could be rewritten as

$$\ddot{\varphi}_1 = -UJ(\varphi_1 - \varphi_2) + 2UJ'' \cos(\varphi_1) - \gamma\dot{\varphi}_1 + F(t) \qquad (S6)$$
$$\ddot{\varphi}_2 = UJ(\varphi_1 - \varphi_2) - 2UJ'' \cos(\varphi_2) - \gamma\dot{\varphi}_2 + F(t) \qquad (S7)$$

where $\gamma$ is the damping constant and $F(t)$ is a uniform external driving. For an applied electric field $E(t) = E_0 \sin(\omega_{pump} t)$, $F(t) = \dot{E}(t) \sim E_0 \cos(\omega_{pump} t)$

The energy conserving part of the equations can be derived from an effective potential,

$$V_{eff}(\varphi_1, \varphi_2) = \frac{UJ}{2}(\varphi_1 - \varphi_2)^2 + UJ'' \sin(\varphi_1) - UJ'' \sin(\varphi_2) \qquad (S8)$$

The first term in Eq. (S8) represents the in-plane elastic energy while the other terms represent the frustrated inter-layer Josephson energy. These terms makes the effective potential slightly curved (Fig. S6) and this result in two important features.

1) The equilibrium state is one with the phase at each lattice site has a small tilt angle $\left(\varphi_1 = -\varphi_2 \approx \frac{J''}{J}\right)$ on top of the collinear configuration.

2) In the presence of an external applied field $E(t)$, $\varphi_1$ and $\varphi_2$ is driven in-phase by $F(t) = \dot{E}(t)$, while the curved shape of the potential $V_{eff}$ induces small out-phase motions. Such dynamics therefore confirms the existence of two collective modes $\varphi_0 = (\varphi_1 + \varphi_2)/2$ and $\varphi_\pi = (\varphi_1 - \varphi_2)/2$.

Assuming that $\varphi_0 \ll \varphi_\pi$, one could rewrite Eq. (S6) and Eq. (S7) as:

$$\ddot{\varphi}_0 = -2UJ'' \sin(\varphi_0) \sin(\varphi_\pi) - \gamma\dot{\varphi}_1 + F(t) \approx -2UJ'' \varphi_0 \varphi_\pi - \gamma\dot{\varphi}_0 + F(t), \qquad (S9)$$
$$\ddot{\varphi}_\pi = 2UJ'' \cos(\varphi_0) \cos(\varphi_\pi) - 2UJ\varphi_\pi - \gamma\dot{\varphi}_1 \approx 2UJ'' - 2UJ\varphi_\pi - UJ'' \varphi_0^2 - \gamma\dot{\varphi}_\pi. \qquad (S10)$$

It is evident from Eq. (S9) that the applied field $E(t)$ excites $\varphi_0$. In addition to its response at the excitation frequency $\omega_{pump}$, $\varphi_0$ undergoes plasma oscillation at $\omega_{JP0} = \sqrt{2UJ''^2/J}$. These plasma oscillations arises only when $\varphi_\pi \neq 0$.

Also, it is clear from Eq. (S10) that $\varphi_\pi$ is excited only through $\varphi_0^2$ term, hence $\varphi_\pi$ has a response at $2\omega_{pump}$. Therefore, in the Fourier space, we may approximate $\varphi_0$ by a single mode $\varphi_0(\omega_{pump})$. The $\varphi_\pi$ mode may be approximated with two frequency components $\varphi_\pi(0)$ and $\varphi_\pi(2\omega_{pump})$.

Now the total current is given by

$$I(t) = J'' \cos(\varphi_1) - J'' \cos(\varphi_2) \approx -2J'' \varphi_0(t) \varphi_\pi(t) + \frac{J''}{3} \varphi_0^3(t) \varphi_\pi(t) \tag{S11}$$

In the Fourier space, the same equation reads as:

$$\begin{aligned} I(\omega) \approx &-2J'' \varphi_0(\omega_{pump}) \varphi_\pi(0) + \frac{J''}{3} \varphi_0^3(\omega_{pump}) \varphi_\pi(2\omega_{pump}) \\ &-2J'' \varphi_0(\omega_{pump}) \varphi_\pi(2\omega_{pump}) + \frac{J''}{3} \varphi_0^3(\omega_{pump}) \varphi_\pi(0) \end{aligned} \tag{S12}$$

The contributions to the third harmonic generation (current response at $3\omega_{THz}$) could be divided into two terms:

1) A term proportional to the $\varphi_0^3(\omega_{pump}) \varphi_\pi(0)$. This term can exist only when $\varphi_\pi \neq 0$. Or in other words, when there is finite superfluid tunneling between the layers which occurs at $T < T_c$.

2) The term proportional to $\varphi_0(\omega_{pump}) \varphi_\pi(2\omega_{pump})$ which arises from the nonlinear coupled dynamics of the PDW state. Such contributions would be present even when $\varphi_\pi = 0$ and hence could be postulated to explain the experimental observation of third harmonic generation (THG) at $T > T_c$ in the stripe ordered state.

We have numerically integrated Eq. (S6) and (S7) with an initial steady state until $t = 300$. Note that the time coordinate was normalized with $\omega_{pump}$. The computations were performed with $J = 4$, $J'' = 0.5$, $U = 1$, $\gamma = 0.2$ and $\omega_{THz} = 1$. The parameters were chosen so as to match the experiments ($\omega_{JP0} = 0.15 \, THz$, $\omega_{THz} = 0.45 \, THz$, $\frac{\omega_{JP0}}{\omega_{pump}} = 0.35$). The driving is taken as $F(t) = F_0 \exp(-\frac{(t-t_E)^2}{2\Delta_E^2}) \cos(\omega_{pump} t)$ with $t_E = 100$ and $\Delta_E = 5$.

The power spectrum of the current response for driving amplitudes of $F_0 = 0.05$ and $F_0 = 0.5$ is shown in in Fig. S7. The strength of the driving amplitude corresponds to tens of kV/cm for $F_0 = 0.5$. A third harmonic response is observed for the larger amplitude of driving. Further, the contribution to the THG from the two terms discussed above is shown in Fig. S7(B). It could be noted that the significant contribution to THG arises from the novel dynamical coupling of the collective modes of the PDW. However, we note the presence of the term proportional to $\varphi_0^3(\omega_{pump}) \varphi_\pi(0)$. This is due to the limitation of our model which does not have a clear phase transition below which the static $\varphi_\pi$ mode (the finite static tilting of the phases) vanishes ($\varphi_\pi = 0$).

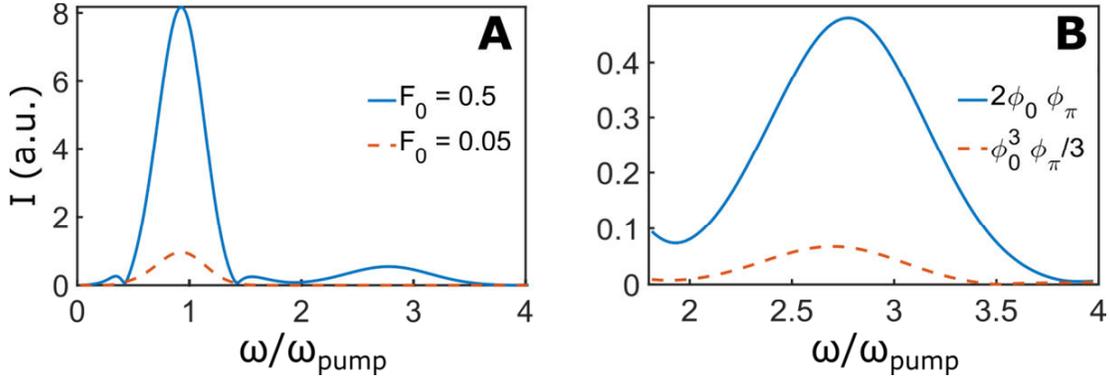

Fig. S7 – (A) The computed current response in the linear ($F_0 = 0.05$) and nonlinear ($F_0 = 0.5$) regimes. A clear response at the third harmonic of the driving field is observed in the nonlinear excitation regime. (B) The contribution of the different terms discussed in the text to the third harmonic intensity.

## S6: Nonlinearities from quasiparticles –First principle DFT calculations

In this section an estimate for a second contribution to the third harmonic We quantify the third harmonic generated from the free carrier motion in the anharmonic Bloch bands. Specifically, the deviation of the band dispersion from a parabola ($\epsilon \approx a_2 k^2 + a_4 k^4$) results in third harmonic emission.

An applied electric field imparts a finite momentum to the free carriers, following which they explore various Bloch states while acquiring group velocities $v_g(t)$ depending on the band curvature. Radiation emerges as the charges are accelerated and hence as

$$\frac{dI_{intra}}{dt} = \frac{d}{dt}[n_e\, v_g(t)] = n_e \frac{dv_g(t)}{dt}, \qquad (S13)$$

with $n_e$ the number of mobile charge carriers within the band and $v_g(t)$ the group velocity.

To obtain the time dependent group velocity we use the known relation involving the time dependent momentum change $k(t)$ and the dispersion relation $\epsilon(k)$ which finally give:

$$v_g(t) = \frac{1}{\hbar} \left.\frac{d\epsilon(k)}{dk}\right|_{k(t)}. \qquad (S14)$$

Lastly, starting from a Drude model the time dependence of the momentum ($k(t)$) is given by

$$\frac{dk(t)}{dt} = \frac{e}{\hbar} E(t) - 2\frac{k(t)}{\gamma} ,  \quad (S15)$$

with *e* the electron charge, E(t) the time dependent electric field and $\gamma$ the mean free time of the electrons.

Experimental and material specific parameters were computed in the following fashion:

1) We computed $k(t)$ with parameters E(t) and mean free time, similar to what is observed in the experiments.
2) We performed computations of the electronic structure of $La_{1.92}Ba_{0.08}CuO_4$ (LBCO) to determine the band structure $\epsilon(k)$.
3) We combined the two steps and perform a Fast Fourier Transformation (FFT) of the resulting $dv_g(t)/dt$, which gives the spectrum of emitted light.

Since the applied THz field is polarized along the *c*-axis of LBCO, the free carriers would accelerate along the $\Gamma - Z$ direction parallel to the reciprocal c* vector (the Brilloun-zone of LBCO is shown in Fig. S8(A)). The k(t) for a single Gaussian pulse with $f = 0.45\ THz$, $E_{max} = 65\ kV/cm$, $FWHM = 3\ ps$ and $\gamma = 70\ fs$ was computed from Eq. (S15) (Fig. S8(B)). The resulting momentum oscillations are small with a maximum amplitude not exceeding 10% of c*. Consequently, the electron oscillations are confined close to the center of the Brillouin zone.

We compute the electronic band structure of LBCO by performing first principle computations in the framework of DFT. For our calculations, we use DFT as expanded with an augmented plane wave plus local orbital (APW+lo) basis as implemented in the ELK code(*30*). As approximation for the exchange correlation functional we use the generalized gradient approximation corresponding to Perdue, Burke and Ernzerhof(*31*) (PBE) and to improve the description of correlation effects of the Cu *d*-electrons we employ a DFT+$U$ scheme(*32*). For the latter, we use a U=5eV and J=0 and apply the fully localized limit for the double counting correction term. As a structural input of our computations we use the data provided for LBCO in Ref [(*33*)] and treat the chemical doping by fractional site occupation of La(Ba) using the virtual crystal approximation (VCA). Finally, we perform several convergence tests on the band structure of LBCO and assume convergence if $\Delta\epsilon(k) \leq 0.1\,meV$, which we obtain for the numerical parameters given under Ref. [(34)].

Fig. S8(C) shows the dispersion of the bands in the closest vicinity of the Fermi level, two valence and one conduction band, along the $\Gamma - Z$ direction within the Brillouin zone. Please note, the large separation of the valence and conduction band in the order of $2\,eV$, which is three order of magnitude larger compared to the THz pulse frequency. Consequently, the amount of carries generated by THz field in these bands is expected to be small.

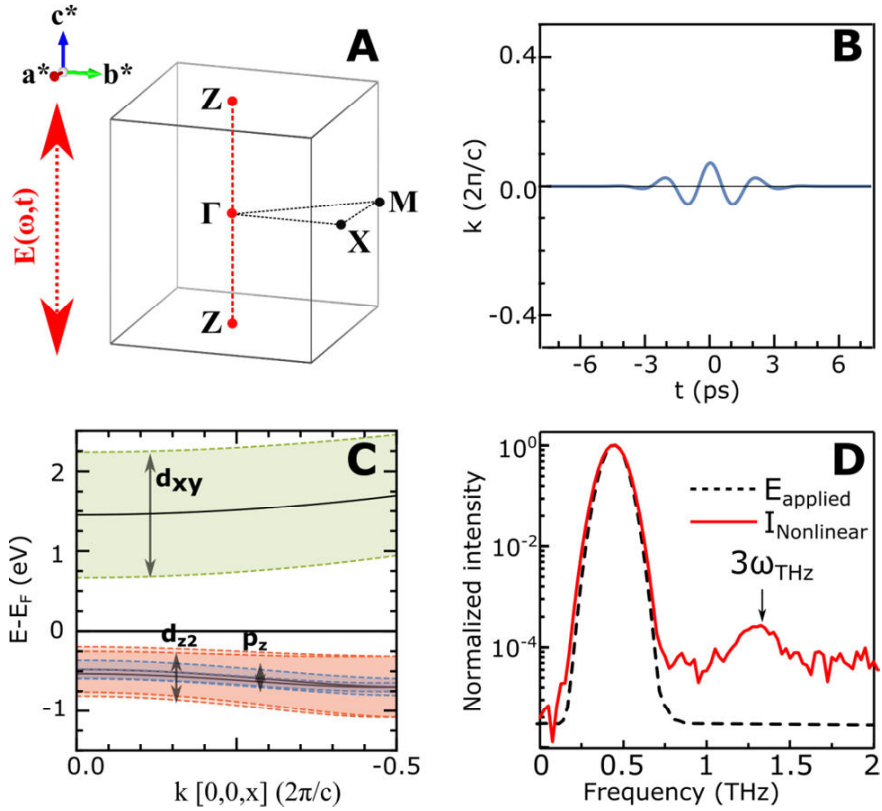

Figure S8 – (A) Brillouin Zone of LBCO x=9.5% doped sample indicating the high symmetry points. (B) The $k(t)$ excited along the Γ-Z direction by an electric field E(t), which frequency spectra is shown in (D). (C) The electronic band structure along Γ-Z direction indicating three bands (one conduction and two valence bands). The fat band color coding indicates the weighted symmetry of the bands. (D) The computed current response and applied electric pulse. The former clearly indicates a weak third harmonic contribution.

We also analyze the symmetry of the bands plot the band characters by as fatbands plot in Fig. S8(C). The character of conduction band is $d_{xy}$ whereas the valence bands exhibit a combination of $d_{z^2} + p_z$. Considering, in this respect the dipole transition matrix elements between these bands we find them vanishing since the combination of symmetries does not obey the dipole transition rules ($\Delta l = 1, \Delta m = 0, \pm 1$). Consequently, beside the large energy separation also band character strongly suppresses transitions and the generation of movable charge carriers. Note that also indirect tunneling transitions, which are according Ref. [(*21*)] given by the square of the dipole transition matrix elements, become strongly suppressed. Consequently, even without explicitly computing the quantitate amount of carriers we anticipate a negligible amount of total emitted harmonics.

Independent of the only small amount of available charge carriers we compute the relative magnitude of the third harmonic generated light by evaluating Eq. (S13) after fitting the dispersion relation with a polynomial expression. We further sum the contribution of from the three bands considering the same electron/hole content in each. The resulting emitted light spectrum after the FFT is displayed in Fig. S8(D). We find a strong component at the fundamental, whereas the third harmonic component is 3-4 orders of magnitude smaller. Please, note that increasing the amount of charge carriers will leave the relative size of the generated light unchanged since $n_e$ equally scales all components independent of the frequency.

**S7: Nonlinearities from charge order – Negligible contribution**

In this section we present our argument against the charge density wave (CDW) as the cause of the observed nonlinearities. Although de-pinning a charge density wave can, in principle, generate radiation at the Third-Harmonic (*22*), we show here that it is highly unlikely under the experimental conditions.

The frequency dependence of the cubic dielectric constant for a sliding CDW scales as - $\epsilon^{(3)} \sim E_{inc}^2 \omega_{inc}^{-5}$, where $E_{inc}$ and $\omega_{inc}$ are the field strength and the frequency of the applied electric field (*22*). Starting from this formula, even considering the larger fields used in our experiment (100 kV/cm in the present work vs V/cm in Ref. (*22*)) and considering the appropriate frequency difference (0.5 THz in the present work vs 1 Hz - 8 KHz in Ref. (*22*)), one finds that the Third Harmonic from a CDW should be atleast thirteen orders of magnitude weaker in our experiment.

Physically, this is understood by considering that the total kinetic energy acquired by a confined particle in an electromagnetic field scales with $1/\omega^2$. Hence, the velocity of a charged particle is between sixteen and twenty orders of magnitude larger at 10 KHz and 1Hz than at 1 THz. The lower energy results in a smaller cubic anharmonicities and explains the difference discussed above.